# Binding energy of a Cooper pairs with non-zero center of mass momentum in d-wave superconductors

M.V. Eremin and I.E. Lyubin
*Kazan State University, Kremlevskaya, 18, Kazan 420008, Russian Federation*
*E-mail: igor606@rambler.ru*

**PACS**: 74.20.-z; 74.60.Jg

**Keywords**: High-$T_c$ superconductors, depairing current

The binding energy of Cooper pairs has been calculated for the case of d-wave symmetry of the superconducting gap in layered cuprate superconductors. We assume that Cooper pairs are formed by the short range potential and then derive the binding energy in the form $\Delta_{\mathbf{kq}} = \Delta_x(\mathbf{q})\cos k_x a + \Delta_y(\mathbf{q})\cos k_y a + \Omega_x(\mathbf{q})\sin k_x a + \Omega_y(\mathbf{q})\sin k_y a$, where $\mathbf{q}$ is a total momentum of the pair. Numerical solutions of the self-consistent system of the integral equations for quantities $\Delta_x(\mathbf{q})$, $\Delta_y(\mathbf{q})$ and $\Omega_x(\mathbf{q})$, $\Omega_y(\mathbf{q})$ along different lines in $q_x$, $q_y$ plane have been obtained. Anisotropy of the depairing total momentum (or depairing current) has been calculated.

## 1. Introduction

The problem of the binding energy of a Cooper pair in the BCS model (phonon pairing mechanism) was discussed by Cooper in his pioneering paper [1] and later on by Casas et al. [2]. It was founded that $\Delta_q = \Delta_0 - cq$, i.e. Cooper pairs are destroyed very fast when they start to move. This behavior of Cooper pairs is in strong contrast to usual bosons. The critical value of momentum at which Cooper pairs are self destroyed can be called as depairing momentum. In *d*-wave HTSC superconductors the energy gap is given by $\Delta_{\mathbf{k}\,\mathbf{q}=0} = \Delta_0(\cos k_x a - \cos k_y a)/2$. As one can see the gap is zero along the diagonals in the Brillouin zone, i.e. when $k_x = \pm k_y$. Therefore, one can expect that depairing momentum (or depairing current) in *d*-wave superconductors should be also strongly anisotropic. To our knowledge this problem was not discussed, yet. The goal of present letter is to cover this field. In section 2 we shall derive the integral equation for the order parameter (or for simple explanation – the binding energy), using Hubbard projection technique. The results of numerical solution, derived equation, will be given in section 3.

## 2. Model and Integral Equations

We start from the so-called singlet correlated band model for layered hole–doped cuprates [3]:

$$H = \sum_{ij\sigma} t_{ij} \psi_i^{pd,\sigma} \psi_j^{\sigma,pd} - \frac{1}{2}\sum_{i,j} J_{ij}\left(\psi_i^{\uparrow,\uparrow}\psi_j^{\downarrow,\downarrow} + \psi_i^{\downarrow,\downarrow}\psi_j^{\uparrow,\uparrow} - \psi_i^{\downarrow,\uparrow}\psi_j^{\uparrow,\downarrow} - \psi_i^{\uparrow,\downarrow}\psi_j^{\downarrow,\uparrow}\right) + \frac{1}{2}\sum_{ij} G_{ij}\left(\psi_i^{pd,pd} - \psi_i^{0,0}\right)\left(\psi_j^{pd,pd} - \psi_j^{0,0}\right), \quad (1)$$

where the first term is a kinetic energy, the second is superexchange interaction and the third is density-density interaction, or pseudo-Coulomb interaction. We prefer to speak about the density-density interaction, because the indirect interaction quasiparticles via optical phonons can be taken into account via renormalization of parameter $G_{ij}$. It is assumed that carriers move over oxygen positions in CuO plane. Strong exchange coupling between oxygen and copper spins has been taken into account during the derivation of Hamiltonian (1) [see Ref. 3]. Thus, the symbol $pd$ corresponds to cooper-oxygen singlet state. The equations of motion are given by

$$i\hbar\frac{\partial \psi_i^{\downarrow,pd}}{\partial t} = \left[\psi_i^{\downarrow,pd}, H\right] = \sum_j E_{ij}\psi_j^{\downarrow,pd} + \sum_j D_{ij}\psi_j^{pd,\uparrow}$$

$$i\hbar\frac{\partial \psi_i^{pd,\uparrow}}{\partial t} = \left[\psi_i^{pd,\uparrow}, H\right] = \sum_j E_{ij}\psi_j^{pd,\uparrow} + \sum_j D_{ij}\psi_j^{\downarrow,pd}$$
. (2)

The coefficient $E_{ij}$ and $D_{ij}$ are calculated using following equality [4, 5]:

$$\left\langle\left\{\left[\psi_l^{\downarrow,pd}, H\right], \psi_i^{pd,\downarrow}\right\}\right\rangle = E_{li}\left\langle\left\{\psi_l^{\downarrow,pd}, \psi_i^{pd,\downarrow}\right\}\right\rangle$$

$$\left\langle\left\{\left[\psi_l^{\downarrow,pd}, H\right], \psi_i^{\uparrow,pd}\right\}\right\rangle = D_{li}\left\langle\left\{\psi_l^{pd,\uparrow}, \psi_i^{\uparrow,pd}\right\}\right\rangle,$$

where the angular brackets denote the thermodynamic averaging.
After calculation of the anticommutators in both side, in particular, we have obtained

$$D_{li} = \frac{\delta_{il}}{P_i^\uparrow}\sum_j t_{lj}\left\langle\psi_l^{\uparrow,pd}\psi_j^{\downarrow,pd} - \psi_l^{\downarrow,pd}\psi_j^{\uparrow,pd}\right\rangle + \frac{J_{li}}{P_i^\uparrow}\left\langle\psi_l^{\downarrow,pd}\psi_i^{\uparrow,pd} - \psi_l^{\uparrow,pd}\psi_i^{\downarrow,pd}\right\rangle - \frac{G_{li}}{P_i^\uparrow}\left\langle\psi_l^{\downarrow,pd}\psi_i^{\uparrow,pd}\right\rangle. \quad (3)$$

Here is $P_i^\uparrow = \left\langle\psi_i^{\uparrow,\uparrow} + \psi_i^{pd,pd}\right\rangle$. For *d*-wave superconductors the sum in (3) is zero, so we can drop it. Performing the Fourier transform in Eq. (3) in the usual way

$$\psi_i^{\downarrow,pd} = \frac{1}{\sqrt{N}}\sum_\mathbf{k}\psi_\mathbf{k}^{\downarrow,pd} e^{i\mathbf{k}\mathbf{R}_i}, \quad (4)$$

and introducing the total momentum of the pair $\mathbf{q} = \mathbf{k}_1 + \mathbf{k}_2$, we get:

$$i\hbar\frac{\partial \psi_\mathbf{k}^{\downarrow,pd}}{\partial t} = \varepsilon_\mathbf{k}\psi_\mathbf{k}^{\downarrow,pd} + \Delta_{\mathbf{k}\mathbf{q}}\psi_{-\mathbf{k}+\mathbf{q}}^{pd,\uparrow}$$

$$i\hbar\frac{\partial \psi_{-\mathbf{k}+\mathbf{q}}^{pd,\uparrow}}{\partial t} = -\varepsilon_{-\mathbf{k}+\mathbf{q}}\psi_{-\mathbf{k}+\mathbf{q}}^{pd,\uparrow} + \Delta_{\mathbf{k}\mathbf{q}}^*\psi_\mathbf{k}^{\downarrow,pd}$$
.

The order parameter is determined as follows:

$$\Delta_{\mathbf{k}\mathbf{q}} = \frac{1}{NP^\uparrow}\sum_{\mathbf{k}_1}\left(J(\mathbf{k}_1 - \mathbf{k}) + J(\mathbf{q} - \mathbf{k}_1 - \mathbf{k}) - G(\mathbf{k}_1 - \mathbf{k})\right)\left\langle\psi_{\mathbf{k}_1}^{\downarrow,pd}\psi_{-\mathbf{k}_1+\mathbf{q}}^{\uparrow,pd}\right\rangle. \quad (5)$$

It is interesting to note the exchange part of the kernel in Eq. (5), namely, $J(\mathbf{q}-\mathbf{k}_1-\mathbf{k})$ contains the total momentum $\mathbf{q}$. This interesting feature, was not pointed out before, because the most part of investigations in this field were done taking into account the density-density interaction only.

Then, we applied Green's function method and completed the system of the equations

$$(\varepsilon_{\mathbf{k}} - E)\langle\langle\psi_{\mathbf{k}}^{\downarrow,pd}|\psi_{\mathbf{k}}^{pd,\downarrow}\rangle\rangle_E + \Delta_{\mathbf{kq}}\langle\langle\psi_{-\mathbf{k}+\mathbf{q}}^{pd,\uparrow}|\psi_{\mathbf{k}}^{pd,\downarrow}\rangle\rangle_E = -\frac{i}{2\pi}P^{\downarrow}$$

$$\Delta_{\mathbf{kq}}^*\langle\langle\psi_{\mathbf{k}}^{\downarrow,pd}|\psi_{\mathbf{k}}^{pd,\downarrow}\rangle\rangle_E + (-\varepsilon_{-\mathbf{k}+\mathbf{q}} - E)\langle\langle\psi_{-\mathbf{k}+\mathbf{q}}^{pd,\uparrow}|\psi_{\mathbf{k}}^{pd,\downarrow}\rangle\rangle_E = 0$$

Solving this system we get the energy of Bogoliubov's quasiparticles spectrum:

$$E_{1,2} = \frac{\varepsilon_{\mathbf{k}} - \varepsilon_{-\mathbf{k}+\mathbf{q}}}{2} \pm \sqrt{\frac{1}{4}(\varepsilon_{-\mathbf{k}+\mathbf{q}} + \varepsilon_{\mathbf{k}})^2 + |\Delta_{\mathbf{kq}}|^2} . \tag{6}$$

Further, using the well known formula

$$\langle\psi_{\mathbf{k}}^{pd,\downarrow}\psi_{-\mathbf{k}+\mathbf{q}}^{pd,\uparrow}\rangle = \int_{-\infty}^{\infty}\frac{dE}{e^{\beta E}+1}\lim_{\varepsilon\to 0}\left(\langle\langle\psi_{-\mathbf{k}+\mathbf{q}}^{pd,\uparrow}|\psi_{\mathbf{k}}^{pd,\downarrow}\rangle\rangle_{E+i\varepsilon} - \langle\langle\psi_{-\mathbf{k}+\mathbf{q}}^{pd,\uparrow}|\psi_{\mathbf{k}}^{pd,\downarrow}\rangle\rangle_{E-i\varepsilon}\right),$$

we find the correlation function:

$$\langle\psi_{\mathbf{k}}^{pd,\downarrow}\psi_{-\mathbf{k}+\mathbf{q}}^{pd,\uparrow}\rangle = -P^{\downarrow}\frac{\Delta_{\mathbf{kq}}^*}{E_1 - E_2}\left(\frac{1}{e^{E_1\beta}+1} - \frac{1}{e^{E_2\beta}+1}\right) = -P^{\downarrow}\frac{\Delta_{\mathbf{kq}}^*}{E_1 - E_2}(f(E_1) - f(E_2)),$$

where $f(E) = \frac{1}{e^{E\beta}+1}$ is Fermi function. Substituting the correlation function in the equation (5) and we get the integral equation, which must be solved self-consistently

$$\Delta_{\mathbf{kq}} = -\frac{1}{N}\sum_{\mathbf{k}_1}(J(\mathbf{k}_1-\mathbf{k}) + J(\mathbf{q}-\mathbf{k}_1-\mathbf{k}) - G(\mathbf{k}_1-\mathbf{k}))\frac{\Delta_{\mathbf{kq}}}{E_1 - E_2}(f(E_1) - f(E_2)). \tag{7}$$

The energy dispersion of quasiparticles in hole-doped HTSC is written as

$$\varepsilon_{\mathbf{k}} = 2t_1(\cos k_x a + \cos k_y a) + 4t_2 \cos k_x a \cos k_y a + ... \tag{8}$$

where hopping integrals $t_1$, $t_2$ ... and chemical potential $\mu$ were taken in accordance with Norman's analysis [6], who extracted all these parameters from the photoemission and neutron scattering data. The Fourier-transform of the exchange potential is given by:

$$J(\mathbf{k}) = J_1(\cos k_x a + \cos k_y a). \tag{9}$$

Note, this form of Fourier-transform is valid for any short-range interaction. As an example one may assume that its role plays the superexchange interaction between of copper spin, then $J_1 \approx 100$ meV. The solution of equation (7), for the case $q = 0$ was described earlier [7]. Below we shall assume that the exchange interaction dominates. Therefore, the Coulomb-like term in equation (7) will be dropped.

### 3. Numerical results and Conclusions

Equation (7) belongs to the class of the separable integral equations. It is easy to see that its general solution may be written in the form:

$$\Delta_{\mathbf{kq}} = \Delta_x(\mathbf{q})\cos k_x a + \Delta_y(\mathbf{q})\cos k_y a + \Omega_x(\mathbf{q})\sin k_x a + \Omega_y(\mathbf{q})\sin k_y a . \tag{10}$$

After substitution (10) in the equation (7) we arrive to the system of equations:

$$\Delta_x(\mathbf{q}) = \frac{1}{N}\sum_{\mathbf{k}_1}(J_1\cos k_{1x}a + J_1\cos(q_x - k_{1x})a)\frac{\Delta_{\mathbf{kq}}}{E_1 - E_2}(f(E_2) - f(E_1))$$

$$\Delta_y(\mathbf{q}) = \frac{1}{N}\sum_{\mathbf{k}_1}(J_1\cos k_{1y}a + J_1\cos(q_y - k_{1y})a)\frac{\Delta_{\mathbf{kq}}}{E_1 - E_2}(f(E_2) - f(E_1))$$

$$\Omega_x(\mathbf{q}) = \frac{1}{N}\sum_{\mathbf{k}_1}(J_1\sin k_{1x}a + J_1\sin(q_x - k_{1x})a)\frac{\Delta_{\mathbf{kq}}}{E_1 - E_2}(f(E_2) - f(E_1))$$

$$\Omega_y(\mathbf{q}) = \frac{1}{N}\sum_{\mathbf{k}_1}(J_1\sin k_{1y}a + J_1\sin(q_x - k_{1x})a)\frac{\Delta_{\mathbf{kq}}}{E_1 - E_2}(f(E_2) - f(E_1))$$

$$\tag{11}$$

The system was solved numerically. Resulting dependencies of $\Delta_x(\mathbf{q})$, $\Delta_y(\mathbf{q})$ and $\Omega_x(\mathbf{q})$, $\Omega_y(\mathbf{q})$ on $qa$, where $a$ is the lattice constant, are shown in Fig. 1 and Fig. 2, where, for short, we have dropped the symbol $(\mathbf{q})$.

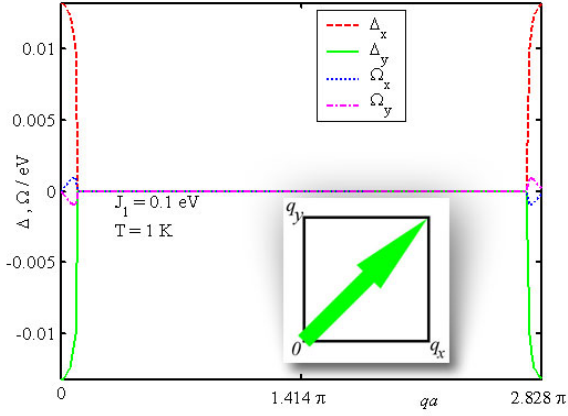

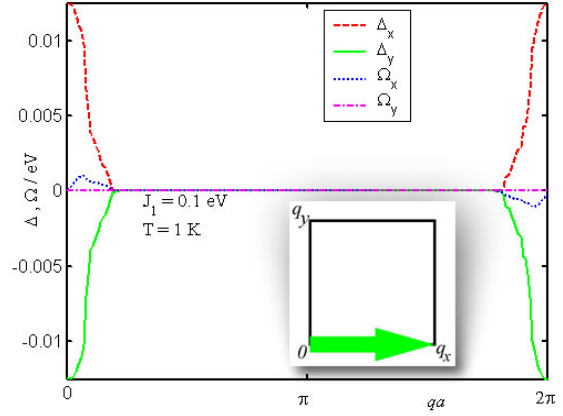

**Fig.1.** Dependences $\Delta_x$, $\Delta_y$, $\Omega_x$, $\Omega_y$ on $qa$ along a diagonal of Brillouin zone $(q_x = q_y)$.

**Fig.2.** Dependences $\Delta_x$, $\Delta_y$, $\Omega_x$, $\Omega_y$ on $qa$ along axis $q_x$ at $(q_y = 0)$.

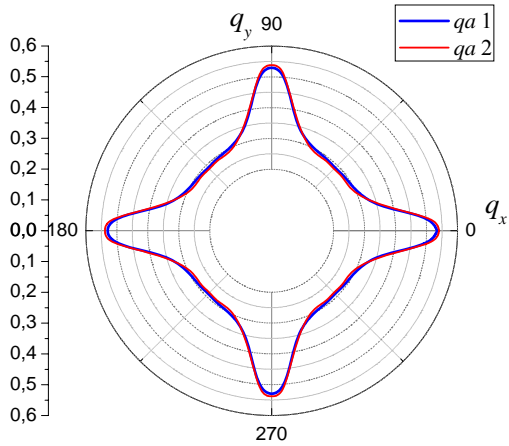

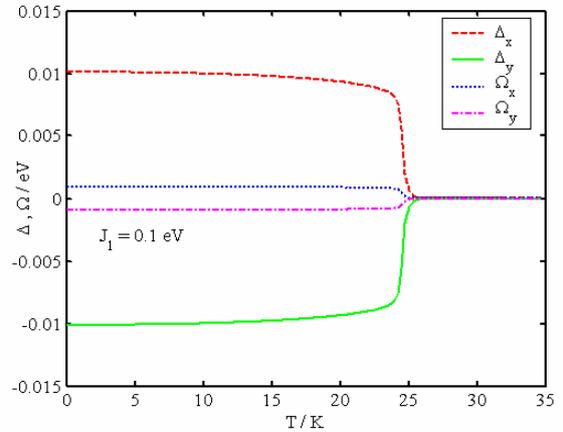

**Fig.3.** Anisotropy of the depairing value $qa$ in $q_x$ and $q_y$ plane.

**Fig.4.** Temperature dependences $\Delta_x$, $\Delta_y$, $\Omega_x$, $\Omega_y$ at q when $|\Omega_x|$ and $|\Omega_y|$ have maximum value.

In Fig. 3 (blue line) we show the anisotropy of the depairing total momentum i.e. the values of $q$ when the binding energy of a Cooper pairs with non-zero center of mass momentum becomes zero. The depairing current can by easily estimated using the simple relation: $j = en\upsilon = en\dfrac{\hbar q}{m_e}$.

From Fig.3 we see that the depairing current in d-wave superconductors in the nodal direction (usual superconducting gap is zero!) approximately twice smaller than in the anti-nodal direction, where the usual gap function has a maximum.

The quantities $\Delta_x(\mathbf{q})$, $\Delta_y(\mathbf{q})$, $\Omega_x(\mathbf{q})$ and $\Omega_y(\mathbf{q})$ /see Eq. (10)/ are monotonically decreasing functions of temperature and all of them vanish at the same critical temperature ($T_c$). In our calculations $T_c = 140K$. In Fig. 4. we show temperature dependences of $\Delta_x(\mathbf{q})$, $\Delta_y(\mathbf{q})$, $\Omega_x(\mathbf{q})$ and $\Omega_y(\mathbf{q})$ at $q$ where $|\Omega_x|$ and $|\Omega_y|$ have maximum value. As one can sees that temperature dependences are quite different from those which are well known for Cooper pairs with zero total momentum.

Let us shortly discuss a variant, when the kernel Eq. (5) does not contain the item like $J(\mathbf{q} - \mathbf{k}_1 - \mathbf{k})$. We note it can be specific just for decoupling procedure which we have used above. In Fig. 3 (red line) we show anisotropy of the depairing total momentum when we have chosen the Fourier–transform interaction in the form $V(\mathbf{k}_1 - \mathbf{k}) = V_0 \left[ \cos(k_1 - k)_x a + \cos(k_1 - k)_y a \right]$ with $V_0 = 200 meV$. As one can see the red and blue curves are very similar, since the depairing momentum is small. Thus, we conclude that presented depairing momentum picture on Fig. 3 is not sensitive to the decoupling procedure.

Finally we would like to point out that if one starts from $t$-$J$ model Hamiltonian instead of (1) (which to our opinion is more relevant to the electron-doped superconductors) then one arrives to the same conclusion as written above, because the kinetic energy item does not enter the integral equation for the order parameter with $d$-wave symmetry.

This work is partially supported by the Swiss National Science Foundation, Grant # IB7420-110784 and the Russian Foundation for Basic Research, Grant # 06-0217197-a.